\documentclass{mpe_report}

\usepackage{psfig,graphicx,epsfig}
\usepackage{color}
\usepackage{amsmath,amssymb,epic,eepic,array}

\begin{document}

\pagenumbering{arabic}
\setcounter{page}{80}

\renewcommand{\FirstPageOfPaper }{ 80}\renewcommand{\LastPageOfPaper }{ 83}

\title{Glitch Observations in Slow Pulsars}
\author{G.H. Janssen\inst{1} \and B.W. Stappers\inst{2}}  
\institute{Astronomical Institute ``Anton Pannekoek'', University of
  Amsterdam, Kruislaan 403, 1098 SJ Amsterdam, The Netherlands
\and  Stichting ASTRON, Postbus 2, 7990 AA Dwingeloo, The Netherlands}
\maketitle

\begin{abstract}
We have analyzed 5.5 years of timing observations of 7 ``slowly''
rotating radio pulsars, made with the Westerbork Synthesis Radio
Telescope. We present improved timing solutions and 30, mostly small
new glitches. The most interesting results are: 1) The detection of
glitches one to two orders of magnitude smaller than ever seen before
in slow radio pulsars. 2) Resolving timing-noise looking structures in
the residuals of PSR\,B1951$+$32 by using a set of small glitches. 3)
The detections of three new glitches in PSR\,J1814$-$1744, a
high-magnetic field pulsar.

In these proceedings we present the most interesting results of
our study. For a full coverage, we refer the reader to Janssen~\&~Stappers~(2006).

\end{abstract}

\section{Glitches: The basics}
Glitches are characterized by a sudden increase of
the pulsar rotation frequency $(\nu)$, accompanied by a change in
spindown rate $(\dot\nu)$ and sometimes followed by relaxation or
exponential decay to the previous rotation state. Typical magnitudes
of glitches are from $10^{-10}\nu$ to $10^{-6}\nu$ and steps in
slowdown rate are on the order of $10^{-3}\dot\nu$.  
Glitches give an unique opportunity to study the internal structure of
neutron stars, as they are believed to be caused by sudden and
irregular transfer of angular momentum from the superfluid inner parts
of the star to the more slowly rotating crust (Ruderman et
al. 1998). They are mostly seen in pulsars with characteristic ages
($\tau_c$) around $10^4-10^5$ yr and can occur up to yearly in some
pulsars.  Most of the youngest pulsars, with $\tau_c\lesssim2000$~yr
show very little glitch activity. This could be because they are still
too hot which allows the transfer of angular momentum to happen more
smoothly (McKenna \& Lyne 1990).

We have observed our sample of pulsars since 1999 at the Westerbork Synthesis
Radio Telescope (WSRT) with the Pulsar Machine (PuMa; Vo\^ute et al. 2002)
at multiple frequencies centered at 328, 382, 840 or 1380 MHz as shown in Table
\ref{tab:frequencies}.

\begin{table}[b]
      \caption{Summary of observed frequencies.}
         \label{tab:frequencies}
      \[
      \begin{tabular}{lcccc}
	    \hline
            \hline
            \noalign{\smallskip}
	    Pulsar Name  &  328 MHz  & 382 MHz  &  840 MHz  &  1380 MHz  \\
            \noalign{\smallskip}
            \hline
            \noalign{\smallskip}
	    B0355$+$54 & $\star$ & $\star$ & $\star$ & $\star$\\
	    B0525$+$21 & $\star$ & $\star$ & $\star$ & $\star$\\
	    B0740$-$28 & $\star$ & $\star$ & $\star$ & $\star$\\
	    B1737$-$30 &  &  & $\star$ & $\star$\\
	    J1814$-$1744 &  &  &  & $\star$\\
	    B1951$+$32 &  &  & $\star$ & $\star$\\
	    B2224$+$65 &  &  & $\star$ & $\star$\\
            \noalign{\smallskip}
             \hline
         \end{tabular}
      \]
   \end{table} 

\begin{figure}
 \centerline{\psfig{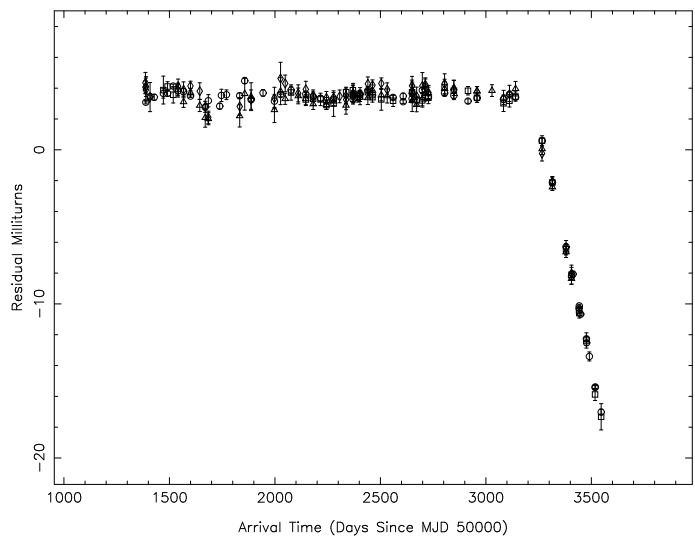} }
 \caption{Residuals for PSR\,B0355+54. A miniglitch is visible at MJD~53200.
 \label{fig:0355}}
\end{figure}

\section{Results: Individual pulsars}
\subsection{PSR\,B0355$+$54}
This relatively old pulsar ($5.6\times 10^5$ yr) has been studied
intensively since its discovery (Manchester et al. 1972).
It has low timing noise and the only report of
glitches in this pulsar is by Lyne (1987) and Shabanova (1990).  The
reported glitches are very different, the first happened at MJD 46079
and was quite small with $\Delta\nu/\nu=5.6\times 10^{-9}$. The
second, at MJD 46496, is one of the largest glitches known, with
$\Delta\nu/\nu=4.4\times 10^{-6}$.

We have a well-sampled multi-frequency observing data span of almost 6
years for this pulsar. We improve on previously published (Hobbs et
al. 2004; Wang et al. 2001) timing solutions. Although a timing
solution for our data span including only position parameters and two
frequency derivatives already yields a better root-mean-square (rms)
of 67~$\mu s$, we find a better solution (47~$\mu s$) including 4
mini-glitches with frequency steps over an order of magnitude smaller
than any glitch found to date in a slowly rotating pulsar:
$\Delta\nu/\nu$ of $10^{-10}$ to $10^{-11}$.  In Fig.~\ref{fig:0355}
one of the small glitches is shown. The residuals show the typical
bend-down signature of a glitch.

\subsection{PSR\,J1814$-$1744}

The estimated surface dipole magnetic field strength of
PSR\,J1814$-$1744 is one of the highest known for radio pulsars,
$5.5\times~10^{13}$G. This is very close to the magnetic field
strengths for anomalous X-ray pulsars (AXPs). The spin parameters are
very similar as well.  However, no X-ray emission was detected for
this pulsar (Pivovaroff et al. 2000).  Three glitches have been
detected in AXPs so far: two in 1RXS~J1708$-$4009 (Dall'Osso et
al. 2003; Kaspi \& Gavriil 2003), and one in 1E~2259$+$586 (Kaspi et
al. 2003). The steps in frequency in AXP glitches seem at least an
order of magnitude larger than those in the high-B radio pulsars, but
the values are not uncommon for radio pulsars in general. For example
the Vela pulsar and PSR\,B1737$-$30 show comparably large-magnitude
glitches.  

\begin{figure}
 \centerline{\psfig{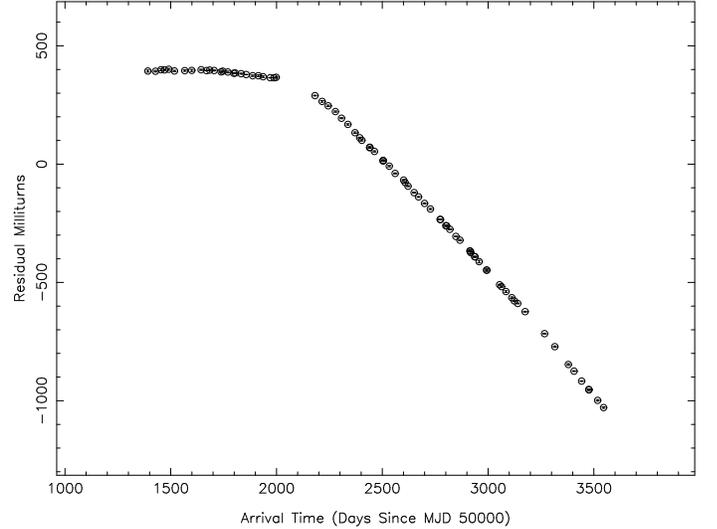} }
 \caption{Glitch detections in the residuals for PSR\,J1814$-$1744. The
  typical glitch signatures are visible aroud MJDs 51700, 52120 and 53300.
\label{fig:1814}}
\end{figure}

We have detected three glitches for PSR\,J1814$-$1744. The
residuals for one of these are shown in Fig.~\ref{fig:1814}.
Together with another glitch detected in the high-magnetic field
pulsar J1119$-$6127 (Camilo et al. 2000), we are now able to compare
AXPs and radio pulsars in another way.

\subsection{PSR\,B1951$+$32}

\begin{figure}
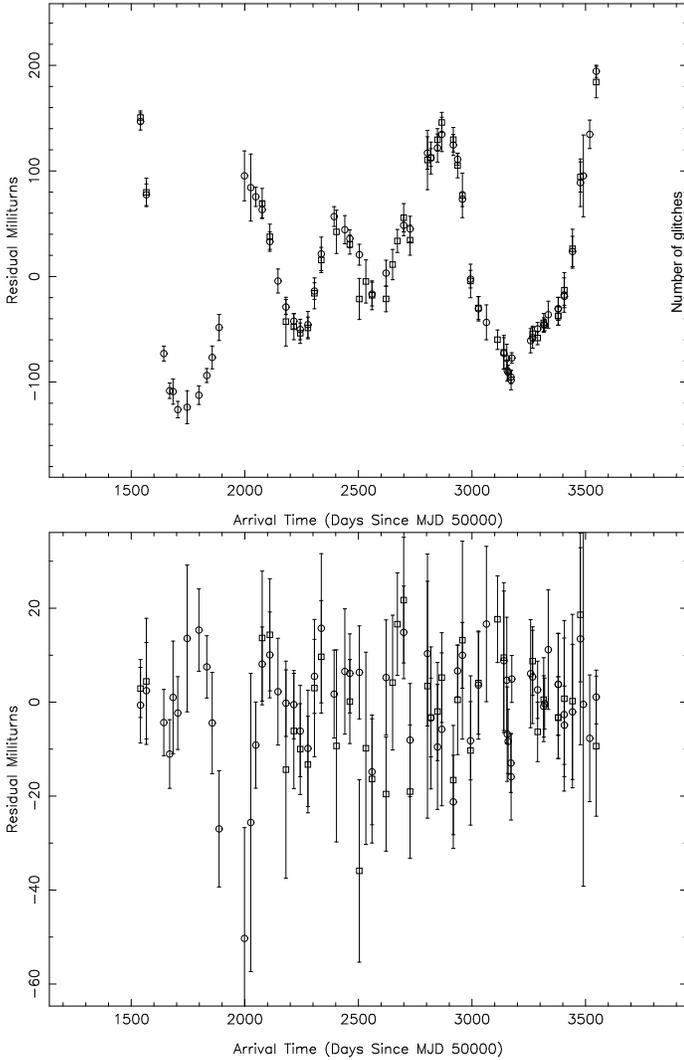

 \centerline{\psfig{file=1951f2chisq1.ps,width=7.0cm,angle=270,clip=} }
 \centerline{\psfig{file=19514glchisq1.ps,width=7.0cm,angle=270,clip=} }
 \caption{Timing residuals for PSR~B1951$+$32. The upper plot shows the
  residuals to a model including two frequency derivatives. The bottom
  plot shows the 10 times better solution including 4 small glitches.
\label{fig:1951}}
\end{figure}

This low-magnetic field pulsar was discovered in 1987 by Kulkarni et
al. (1988), and is associated with the CTB 80 supernova remnant (Strom
1987). A small glitch was detected by Foster et al. (1990) in the
beginning of March 1988.  Because of the high timing activity of this
pulsar, it was quite difficult to find a new timing solution for this
pulsar when starting from a solution with an epoch just before the
start of our data span. Only by shifting the epoch a few hundred days
at a time and creating new solutions for each next epoch could we
generate the present solution with the epoch in the middle of our data
set.  We found a solution for our data span of 5.5 years consisting of
observations at 840 and 1380 MHz. The solution has a rms of 3.2 ms,
which is the best timing solution so far found for this pulsar only
including the first two frequency derivatives. The residuals show a
large timing activity, see the upper plot of Fig.~\ref{fig:1951}.

The pulsar has shown glitching behaviour before, and the cusp-like
structures in the residuals are known to be an indication for glitches
(Hobbs 2002). Therefore we tried to resolve the variations with
glitches. A solution including four glitches results in a much better
rms of 0.4 ms for our data span. The glitches we use are of similar
magnitude to the one reported by Foster et al. (1994), and the steps in
frequency derivative are also comparable. The residuals are shown in
the bottom plot of Fig.~\ref{fig:1951}.

\section{Discussion}

\begin{figure}
 \centerline{\psfig{file=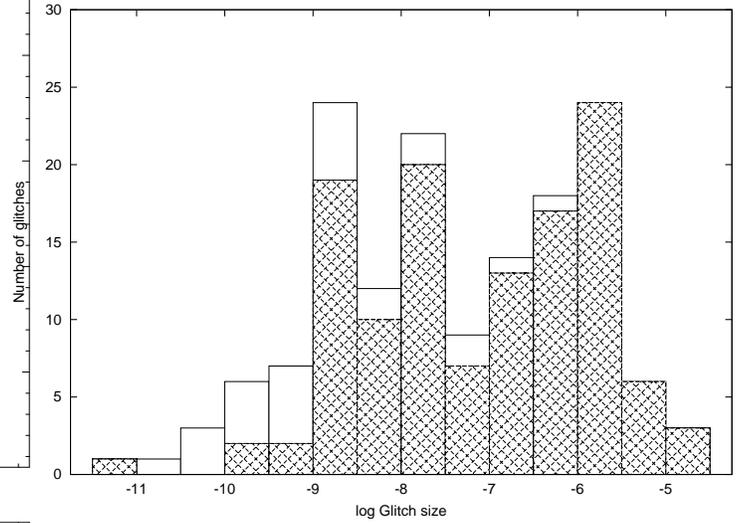,width=7.0cm,angle=270,clip=} }
 \caption{ Histogram of all know glitch sizes, from
  the ATNF glitch table (Manchester et al. 2005) and Urama \& Okeke (1999). 
  New glitches found in this study are added on top of the known glitches.
\label{fig:all-old}}
\end{figure}

\subsection{Glitch sizes}
We have measured glitch sizes down to $\Delta\nu/\nu=10^{-11}$, which
provides the first evidence that such small glitches occur and can be
measured in slowly rotating pulsars.  These glitches are then of a
similar size to the one reported by Cognard \& Backer (2004) in a
millisecond pulsar, and thus perhaps provide further evidence for a
continuous distribution of glitch sizes. Let us now consider how these
small glitches affect the observed glitch size distribution.  In
Fig. \ref{fig:all-old}, a histogram is shown for all now known glitch
sizes.  New glitches found in this study are shown added on top of the
old glitch distribution.  A Kolmogorov-Smirnov test shows that over
the whole range, the distribution has only a probability of $0.001$\%
to be consistent with a flat distribution in log space of glitch
sizes. But if we consider only the part of the diagram between
$10^{-9}<\Delta\nu/\nu<10^{-5.5}$, where the statistics are
better, another KS test shows that the distribution has a $29.8$\%
chance to be drawn from a flat distribution.  The increased number of
glitches with sizes around $\Delta\nu/\nu\approx10^{-9}$, now
comparable to the amount of larger glitches observed, suggests again
that the lack of the smallest glitches at the lower end of the
distribution is due to observing limits.  The lack of glitches at the
upper end of the distribution can not be due to observing
limits. Apparently there is some physical restriction to the maximum
size of a glitch, and we can consider the boundary of
$\Delta\nu/\nu\approx10^{-5}$ as the natural upper limit of glitch
sizes.

\subsection{Glitches vs. timing noise}
Apart from glitches, irregularities in the rotation of the pulsar are
usually described as timing noise. Like glitches, timing noise is also
seen mostly in the younger pulsars with high spin frequency
derivatives.  

To make a better distinction, if possible, between timing noise and
glitches, more modelling is needed, both on the expected glitch size
distributions, as well as on the exact influence on timing parameters
of small glitches and recoveries from large glitches.  We have seen
that for frequently glitching pulsars, it can be difficult to resolve
glitches that occur close together in time.  This effect is probably
more important for small glitches, as they appear to occur more often
and thus are more likely to merge together.  
There are many manifestations of timing noise and some have a form
which clearly cannot be explained as being due to glitches. However our
discovery of small glitches and the way in which we were able to
improve a ''timing-noise-like'' set of residuals for PSR\,B1951$+$32 by
including glitches in the solution indicates that they may play a
role, and that improved sensitivity and more frequent observations may
be required to find more such instances.



          \clearpage

\end{document}